%% file: LCWS11_template.tex
\documentclass[twoside]{LCWS11}
\usepackage{feynmf}
\usepackage[latin1]{inputenc}
\usepackage{graphicx,epsfig,color}
\usepackage{wrapfig,rotating}
\usepackage{amssymb,amsmath,array}
\usepackage{cite}
\usepackage{epstopdf}
\usepackage{url}
\input{symbols}
\begin{document}
\title{
Determination of Top-quark Asymmetries at the ILC} 
\author{Philippe Doublet$^1$, Fran{\c c}ois Richard$^1$, Roman P\"oschl$^1$, Thibault Frisson$^1$, J\'er\'emy Rou\"en\'e$^1$
\thanks{This work is funded by the program {\em Quarks and Leptons} of the IN2P3 France.}
\vspace{.3cm}\\
1- Laboratoire de l'Acc\'{e}l\'{e}rateur Lin\'{e}aire\\
Centre Scientifique d'Orsay, B\^atiment 200\\ 
Universit\'{e} de Paris-Sud XI, CNRS/IN2P3\\ 
F-91898 Orsay Cedex, France
}

\maketitle

\begin{abstract}
A study of top quark production at the future {\em I}nternational {\em L}inear {\em C}ollider, ILC, with a centre-of-mass
energy of 500\,GeV is presented. The emphasis is put on determining the sensitivity to physics beyond the Standard Model. The analysis has been carried out with a full simulation of the ILD detector. Both, the forward-backward asymmetry and the 
left-right asymmetry can be determined to a precision of about 1\% to 1.5\%. The analysis points out an ambiguity which arises in case of the production of top-quarks with left-handed helicity.
\end{abstract}

\section{Introduction}

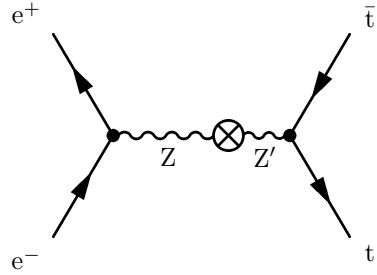
\begin{wrapfigure}{r}{0.5\columnwidth}
\begin{center}
\begin{fmffile}{fmfzzptt}
\begin{fmfchar*}(36,27)\
\fmfleft{em,ep} \fmflabel{$\eplus$}{ep} \fmflabel{$\eminus$}{em}
\fmf{fermion,tension=0.4}{em,Zee,ep}
\fmf{photon}{Zee,v1}
\fmf{photon}{v1,Zff}
\fmfv{decor.shape=circle,decor.filled=empty,decor.size=4mm}{v1}
\fmf{fermion,tension=0.4}{fbar,Zff,f}
\fmfright{f,dum,fbar} \fmflabel{$\tpq$}{f} \fmflabel{$\bar{\tpq}$}{fbar}
\fmfv{decor.shape=cross,decor.size=4mm}{dum}
\fmfdot{Zee,Zff}
\fmffreeze
\fmfshift{0.25w,0.0w}{v1}
\fmfshift{0.3w,0.0w}{Zff}
\fmfshift{0.3w,0.0w}{f}
\fmfshift{0.3w,0.0w}{fbar}
\fmfshift{-0.25w,0.0w}{dum}
\fmfv{label=$\Zzero '$,l.d=0.1w,l.a=-25}{dum}
 \fmfv{label=$\Zzero $,l.d=0.21w,l.a=-157.75}{v1}
\end{fmfchar*}
\end{fmffile}
\end{center}
\caption{\sl Diagram for $\epem \rightarrow \ttbar$ with taking into account new physics which may entail the existence of a $\Zzero '$ boson.}\label{Fig:eett}
\end{wrapfigure}
The top-quark, also $\tpq$-quark hereafter, is the by far heaviest elementary particle known today. With a current world average of $172.0\pm0.6 \pm0.9$~\cite{pdgtop} it is at least about $10^{11}$ times heavier than the lightest fermions. Neither the Standard Model of particle physics nor its super-symmetric extensions deliver an explanation for this striking mass hierarchy. On the other hand, the mass hierarchy can be accommodated in models featuring extra dimensions. The extra degree of freedom w.r.t.\,the Standard Model allows for the arrangement of the particle wave functions along this extra dimensions. An example for these models is that by Randall and Sundrum~\cite{rs99}.  As discussed e.g.\,in~\cite{amr} corresponding effects are more pronounced for heavy quarks. The new physics will modify the electro-weak $\Zzero \ttbar$ vertex and therefore the couplings $\mathrm{Q_L}$ and $\mathrm{Q_R}$ to the left- and right-chiral parts of the $\tpq$-quark wave-function. New physics may also entail the existence of a new $\Zzero '$ boson, see Fig.~\ref{Fig:eett}, or Kaluza-Klein excitations of the Standard Model $\Zzero$-boson. The modified vertex gives rise to different forward-backward asymmetries $\afb$ than those predicted by the Standard Model. The quantity $\afb$ counts events in different detector hemispheres. For this quantity a deviation from the Standard Model expectation has been measured a LEP for $\bottom$-quarks while those for lighter quarks agree with Standard Model expectations~\cite{elwfits}.  The ILC will be operated with polarised electron and positron beams. This in turn permits also to measure the left-right asymmetry $\alr$, i.e. the change in cross-section for different beam polarisations. Here, too, a tension with Standard Model predictions has been observed at SLC~\cite{elwfits}.




\section{Signal selection and background suppression}
This article is based on the studies presented in detail in~\cite{doublet}.  The study assumes a centre-of-mass energy of $\roots=500$\,GeV and  an integrated luminosity of 500\,$\invfb$. The luminosity is shared, where applicable, equally between different polarisations of the incoming beams. The analysis is based on  a full simulation of the detector proposal ILD~\cite{ild09} using event samples generated for the Letter of Intent of the ILC detectors~\cite{ild09, sid-loi}.
The analysis starts out from the process $\epem \rightarrow \ttbar$. The signal process and the main background processes are summarised in Table~\ref{tab:procs}. 
\begin{table}[!h]
\begin{center}
\centerline{\begin{tabular}{|c|c|c|c|c|}
\hline
Channel & $\sigma_{unpol.}$ [fb]& $\sigma_{\eminus_{L}\eplus_{R}}$ [fb]& $\sigma_{\eminus_{R}\eplus_{L}}$ [fb] & $\mathrm{A^{SM}_{LR}}$\%\\ 
\hline
$\tpq \bar{\tpq}$ & 572 & 1564 & 724 & 36.7 \\
\hline
$\mu \mu$ & 456 & 969 & 854 & 6.3 \\
\hline
$\sum_{\mathrm{q=u,d,s,c}} \qq$ & 2208 & 6032 & 2793 & 36.7 \\
\hline
$\bottom \bar{\bottom}$ & 372 & 1212 & 276 & 62.9 \\
\hline
$\mathrm{\gamma} \Zzero$ & 11185 & 25500 & 19126 & 14.2 \\
\hline
$\Wboson \Wboson$ & 6603 & 26000 & 150 & 98.8\\
\hline
$\Zzero \Zzero$ & 422 & 1106 & 582 & 31.0\\
\hline 
$\Zzero \Wboson \Wboson$ & 40 & 151 & 8.7 & 89\\
\hline
$\Zzero \Zzero \Zzero$ & 1.1 & 3.2 & 1.22 & 45\\
\hline
\end{tabular}}
\end{center}
\caption{\sl Unpolarised cross-sections and cross-sections for 100\% beam polarization for signal and background processes. The last column gives the left right asymmetry as expected from the Standard Model.}
\label{tab:procs}
\end{table}%

The produced $\tpq$($\bar{\tpq}$)-quark decays almost exclusively in to a $\bottom \Wboson$ pair. The $\bottom$-quark hadronises giving rise to a jet. The $\Wboson$ can decay {\em hadronically} into light quarks, which turn into jets, or {\em leptonically} into a pair composed by a charged lepton and a neutrino. The {\em semi-leptonic process} is defined by events in which one $\Wboson$ decays hadronically while the other one decays leptonically, i.e.
 \begin{equation}
\ttbar \rightarrow (\bottom \Wboson) (\bottom \Wboson) \rightarrow (\quark \quark') (\bottom \ell \nu) 
\end{equation}
In the Standard Model  the fraction of semi-leptonic final states in  $\epem \rightarrow \ttbar$ is about 43\%. 
The charged lepton, here either $\mathrm{e}$ or $\mathrm{\mu}$ allows for the determination of the top-quark charge. The top quark mass is reconstructed from the hadronically decaying $\Wboson$ which is combined with one of the $\bottom$-quark jets. 

The lepton is either the most energetic particle in a jet which at the same time fulfils typical selection criteria for leptons or has a sizable transverse momentum w.r.t. neighboured jets. Exploiting theses features, the decay lepton can be identified with an efficiency of about 87\%. 

\begin{figure}[ht]
\begin{center}
\includegraphics[width=0.49\textwidth]{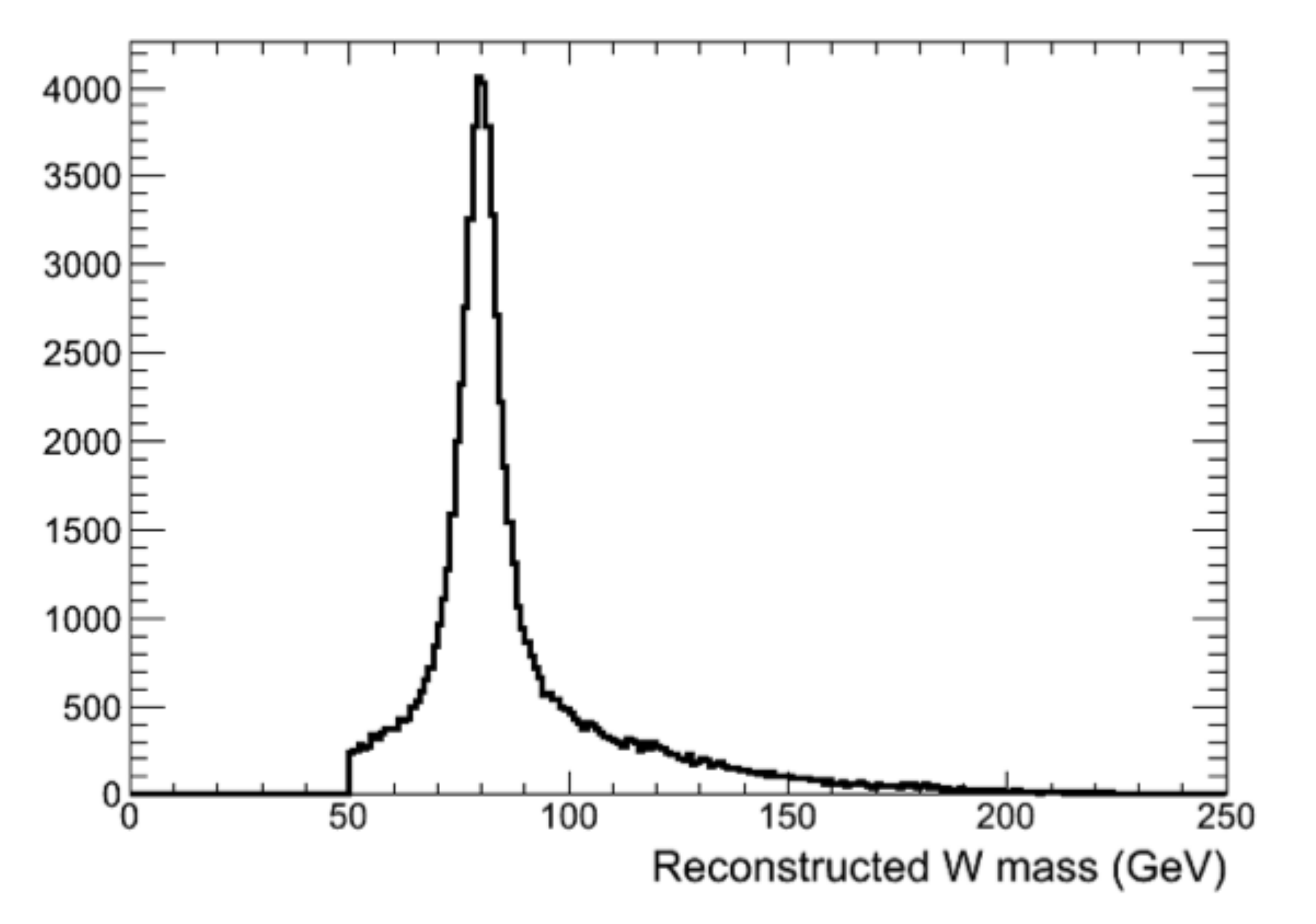}
\includegraphics[width=0.49\textwidth]{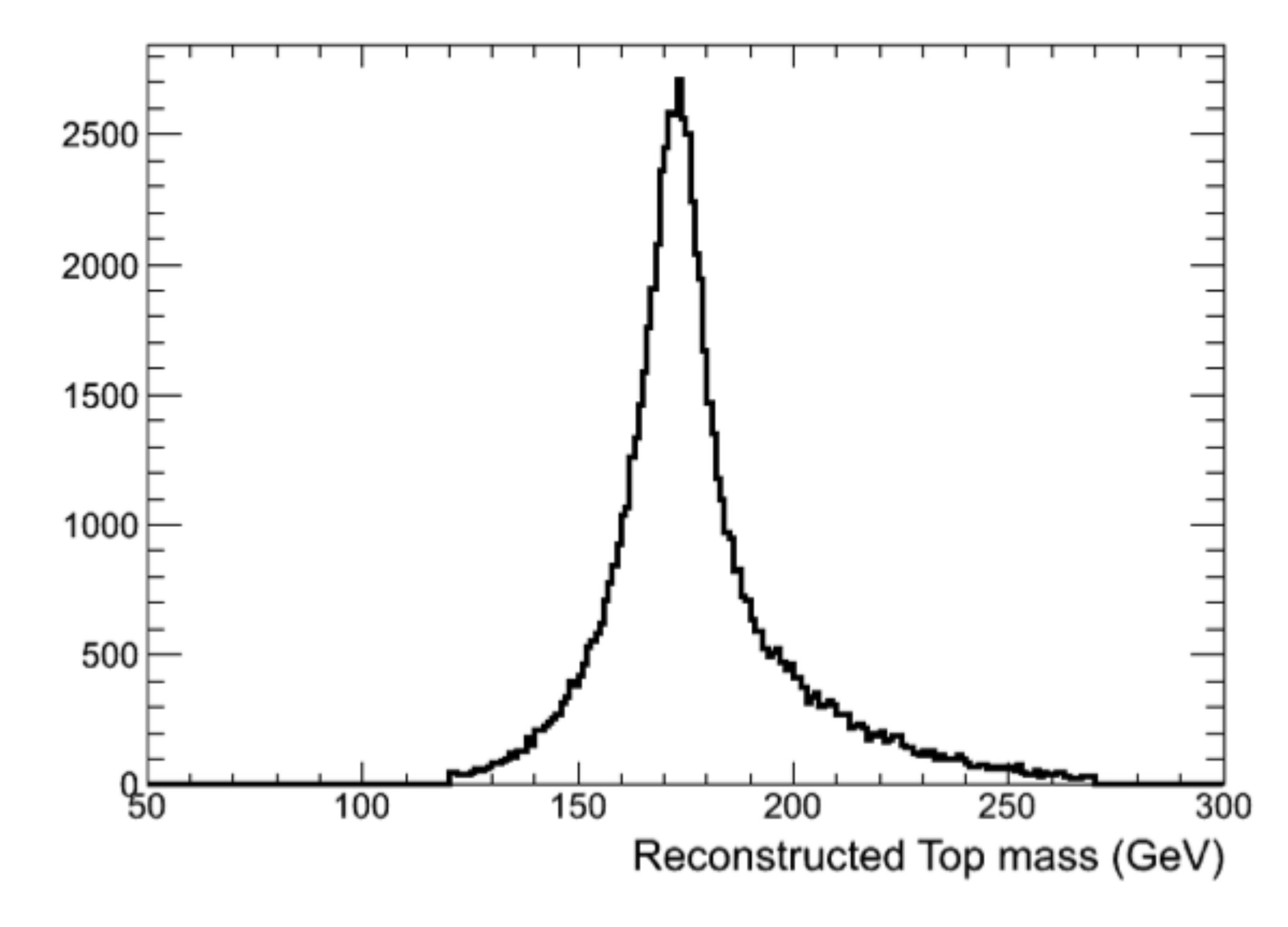}
\caption{\sl Reconstructed masses of the $\Wboson$-quark and the $\tpq$-quark in semi-leptonic decays as obtained in
this analysis.}
\label{fig:tw}
\end{center}
\end{figure}

Of the remaining four jets two must be identified as being produced by the $\bottom$-quarks of the $\tpq$-quark decay. The b-likeness or {\em b-tag} is determined by a neural network which uses information of the tracking system as input. Secondary vertices in the event are analysed by means of the jet mass, the decay length and the particle multiplicity. The jets with the highest b-tag are selected. The b-tag value for the higher one is typically 0.9 while for the smaller one it is still around 0.55. These values are nearly independent of the polar angle of the b-quark jet but drops towards the acceptance limits of the detector. Finally, the two remaining jets are associated with the decay products of the $\Wboson$. The signal is reconstructed by choosing that combination of bottom-quark jet and $\Wboson$ which minimises the following equation:
\begin{equation}
d^2 = (M_{cand.}-M_{\tpq})^2/\sigma^2_{\tpq} + (E_{cand.} - E_{beam})^2/\sigma^2_{E_{\tpq}} + (M^{rec.}_{\Wboson} - M_{\Wboson})^2/\sigma^2_{M_{\Wboson}}
\label{eq:qual}
\end{equation}
 The reconstructed masses of the $\Wboson$ and the top-quark are shown in Figure~\ref{fig:tw}.
A fit of a Gaussian to the maximum of the two spectra results in a resolution of $\sigma_{\Wboson}=4.2$\,GeV for the $\Wboson$-boson mass and of $\sigma_{\tpq}=7.1$\,GeV for the $\tpq$-quark mass. At this stage, the analysis did not target to optimise the precision of these observables but the obtained results indicate that a precision of 50\,MeV or better can be achieved.     
After the selection presented above further cuts against background on jet thrust, hadronic mass of the final state are applied. In addition the mass windows for the reconstructed $\Wboson$-boson and $\tpq$-quark are chosen to $50<m_{\Wboson}<250$\,GeV and $120 < m_{\tpq}<270$\,GeV.
The entire selection retains 72\% signal events. The final sample contains on the other hand 4.6\%  background events.

\section{Results}
The unpolarised cross section for semi-leptonic top-quark decay can be determined to a precision of 0.3\%.  For a pessimistic scenario of electron beams with $\pm 80\%$ polarisation and no positron beam polarisation the resulting statistical precision of the left-right asymmetry $\alr$ is found to be 
\begin{equation}   
\Delta \alr/ \alr = 1.24\%\,(stat.) 
\label{eq:precalr}
\end{equation}

This precision improves considerably in case both beams are polarised as it is foreseen in the baseline of the ILC machine~\cite{rdr07}.

\begin{figure}[!h]
\begin{minipage}[l]{0.45\columnwidth}
\begin{center}
\begin{fmffile}{fmftr}
\begin{fmfgraph*}(36,27)
\fmfleft{i1,i2}\fmflabel{$\bottom$}{i1} 
\fmfright{o1,o2} \fmflabel{$\Wboson$}{o2}
\fmf{plain}{i1,v1,o2}
\fmf{plain}{v1,v2,i1}
\fmf{fermion}{v1,i1}
\fmf{fermion}{v1,o2}
\fmfv{decor.shape=circle, decore.filled=filled,decor.size=2mm}{v1}
\fmfv{label=$\tpq_{\mathrm{R}}$,l.a=0}{v1}
\fmfleft{i3,i4} 
\fmf{dbl_plain_arrow}{i3,i4}
 \fmfv{label=Particle spin,l.d=0.1w,l.a=0}{i3}
\fmfright{o3,o4} 
\fmf{dbl_plain_arrow}{o3,o4}

\fmffreeze
\fmfshift{-0.1w,-0.1w}{v1}
\fmfshift{0.27w,-0.58w}{i4}
\fmfshift{0.1w,-0.0w}{i3}
\fmfshift{-0.25w,0.55w}{o3}
\fmfshift{-0.37w,0.015w}{o4}
\end{fmfgraph*}
\end{fmffile}
\begin{picture}(4,2)
\put(-37,-10){\vector(1,0){10} \raisebox{-0.7mm}{z-Axis}}
\end{picture}
\vspace{20mm}
\begin{fmffile}{fmfkintr}
\begin{fmfgraph*}(36,27)
\fmfleft{bl}
\fmfv{label=$\bottom_{\mathrm{lep.}}$,l.d=-0.2w,l.a=0}{bl}
\fmfright{q1,q3,q2}\fmflabel{$q$}{q1} \fmflabel{$q'$}{q2} \fmflabel{$\bottom_{\mathrm{had.}}$}{q3}
\fmf{fermion}{v1,q1}
\fmf{fermion}{v1,q2}
\fmf{fermion}{v1,bl}
\fmf{fermion}{v1,q3}
\fmffreeze
\fmfshift{-0.4w,0.0w}{q3}
\fmfshift{0.1w,0.0w}{bl}
\fmfshift{-0.1w,-0.1w}{q2}
\fmfshift{-0.1w,0.1w}{q1}
\fmfshift{-0.4w,0.0w}{v1}
\end{fmfgraph*}
\end{fmffile}

\end{center}
\end{minipage}
\hfill
\begin{minipage}[l]{0.45\columnwidth}
\begin{center}
\begin{fmffile}{fmftl}
\begin{fmfgraph*}(36,27)
\fmfleft{i1,i2}\fmflabel{$\Wboson$}{i1} 
\fmfright{o1,o2} \fmflabel{$\bottom$}{o2}
\fmf{plain}{i1,v1,o2}
\fmf{plain}{v1,v2,i1}
\fmf{fermion}{v1,i1}
\fmf{fermion}{v1,o2}
\fmfv{decor.shape=circle, decore.filled=filled,decor.size=2mm}{v1}
\fmfv{label=$\tpq_{\mathrm{L}}$,l.a=0}{v1}
\fmfleft{i3,i4} 
\fmf{dbl_plain_arrow}{i4,i3}
 \fmfv{label=Particle spin,l.d=0.1w,l.a=0}{i3}
\fmfright{o3,o4} 
\fmf{dbl_plain_arrow}{o3,o4}

\fmffreeze
\fmfshift{-0.1w,-0.1w}{v1}
\fmfshift{0.7w,-0.18w}{i4}
\fmfshift{0.53w,0.4w}{i3}
\fmfshift{-0.67w,0.17w}{o3}
\fmfshift{-0.79w,-0.365w}{o4}
\end{fmfgraph*}
\end{fmffile}
\vspace{20mm}
\begin{fmffile}{fmfkintl}
\begin{fmfgraph*}(36,27)
\fmfleft{bl}
\fmflabel{$\bottom_{\mathrm{lep.}}$}{bl}
\fmfright{q1,q3,q2}\fmflabel{$q$}{q1} \fmflabel{$q'$}{q2} \fmflabel{$\bottom_{\mathrm{had.}}$}{q3}
\fmf{fermion}{v1,q1}
\fmf{fermion}{v1,q2}
\fmf{fermion}{v1,bl}
\fmf{fermion}{v1,q3}
\fmffreeze
\fmfshift{0.2w,0.0w}{q3}
\fmfshift{-0.27w,-0.2w}{q2}
\fmfshift{-0.27w,0.2w}{q1}
\fmfshift{-0.07w,0.0w}{v1}
\end{fmfgraph*}
\end{fmffile}
\end{center}
\end{minipage}
\caption{\sl Upper figures: The decay of a $\tpq_{\mathrm{R}}$ quark leads predominantly to a high energetic and longitudinally  polarised $\Wboson$ and a soft $\bottom$ quark. In case of the decay of a $\tpq_{\mathrm{L}}$ the situation is reversed. Lower figures: In case of a $\tpq_{\mathrm{R}}$ decay the jets from the $\Wboson$ can easily associated with the corresponding $\bottom$-quark $\bottom_{\mathrm{had.}}$. In case of a $\tpq_{\mathrm{L}}$ the $\Wboson$ is practically at rest and the association with $\bottom_{\mathrm{had.}}$ and with that of the leptonic decay  $\bottom_{\mathrm{lep.}}$ gives identical results when reconstructing the $\tpq$-mass.}
\label{fig:ambig}
\end{figure}
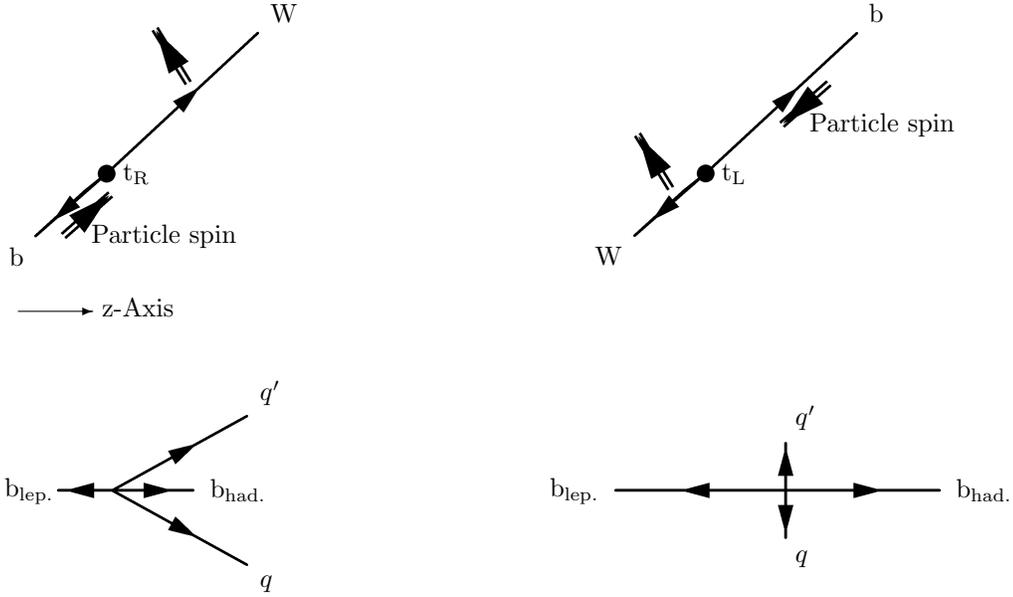

For the determination of the forward-backward asymmetry $\afb$, the number of events in the hemispheres of the detector w.r.t. the polar angle is counted. Again the analysis is carried out separately for a left-handed polarised electron beam and for a right handed polarised beam. In the final state two different situations have to be distinguished, see also Fig.~\ref{fig:ambig}: 
\begin{itemize}
\item In case of a {\em right}-handed electron beam the sample is expected to be enriched with $\tpq$-quarks with {\em right}-handed helicity~\cite{pasha}. Due to the V-A structure of the standard model an energetic $\Wboson$-boson is emitted into the flight direction of the $\tpq$-quark. This $\Wboson$-boson can be easily associated with the corresponding $\bottom$-quark from the hadronic decay of the $\tpq$-quark. Therefore, the polar angle of the $\tpq$-quark can be reconstructed without ambiguities.
\item In case of a {\em left}-handed electron beam the sample is enriched with $\tpq$-quarks with {\em left}-handed helicity. In this case the $\Wboson$-boson is emitted opposite to the flight-direction of the $\tpq$-quark and gains therefore only little kinetic energy. In fact for a centre-of-mass energy of 500\,GeV the $\Wboson$-boson is nearly at rest. In this case the combination with the $\bottom$-quark from the leptonic decay of the $\tpq$-quark and with that from the hadronic decay gives identical results which gives rise to ambiguities in the reconstruction of the $\tpq$-quark. 
\end{itemize}
The explanations above apply correspondingly to polarised positron beams and $\bar{\tpq}$-quarks.

\begin{figure}[ht]
\begin{center}
\includegraphics[width=0.49\textwidth]{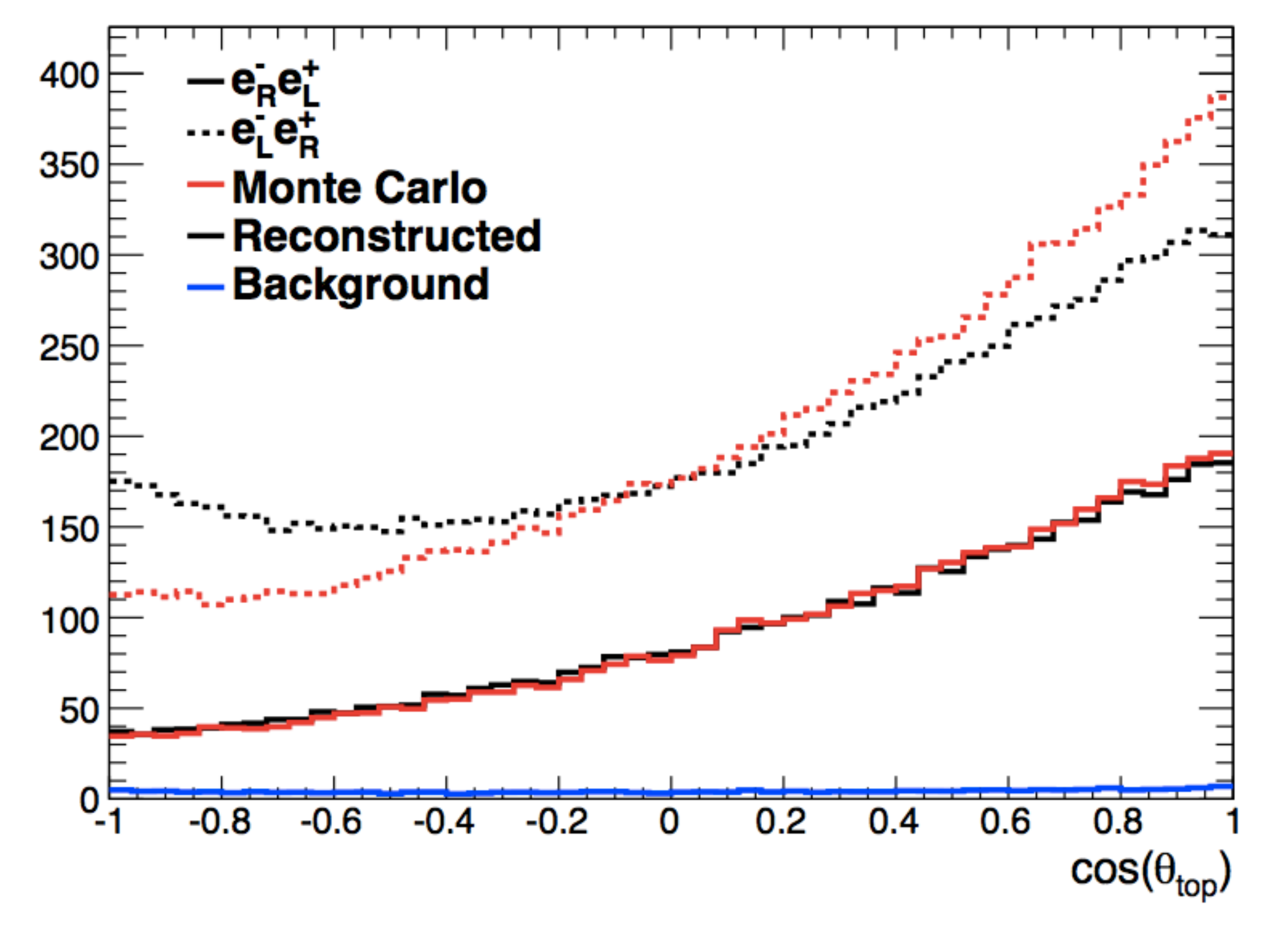}
\includegraphics[width=0.44\textwidth]{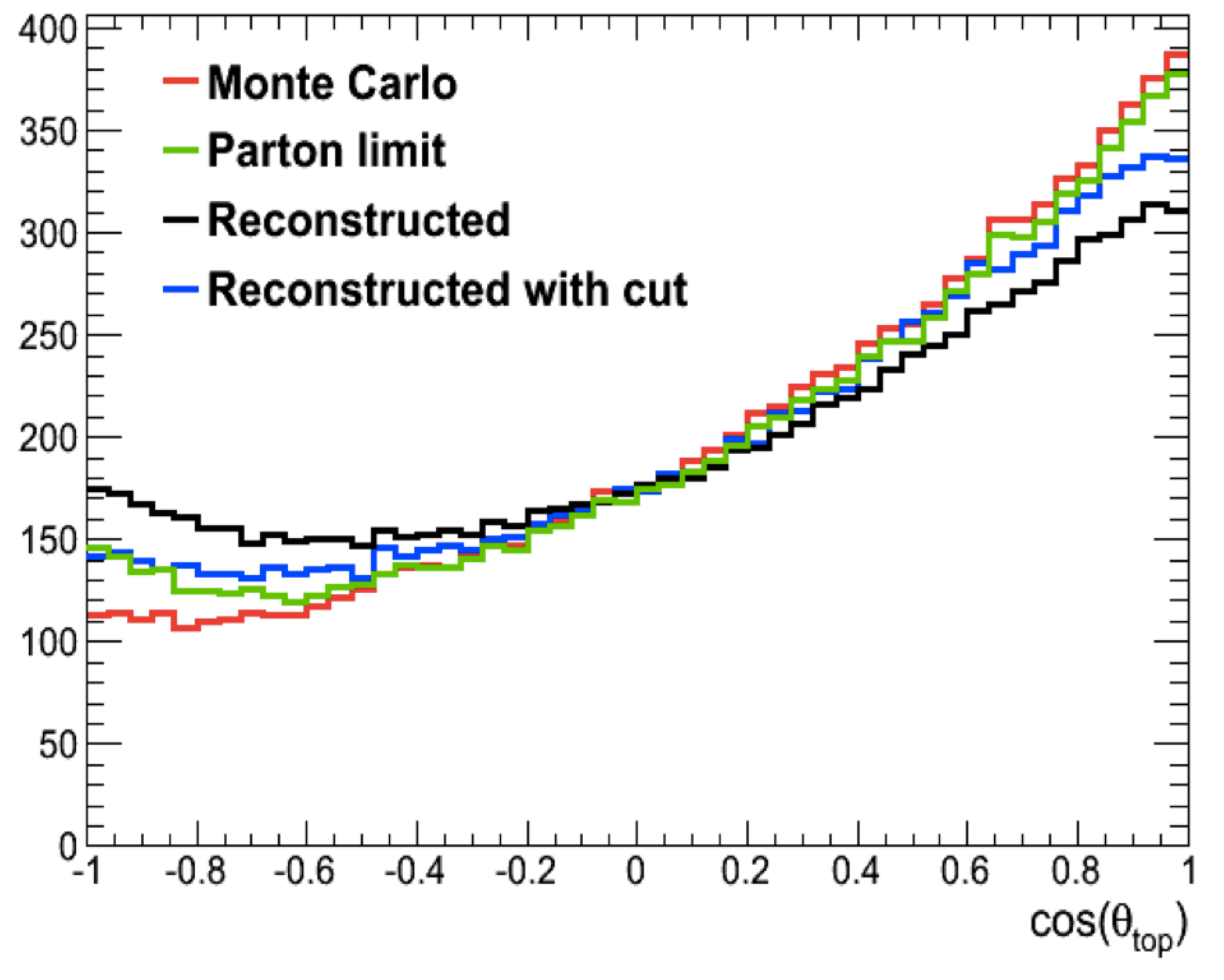}
\caption{\sl Left: Reconstructed polar angle of the $\tpq$-quark for right and left-handed electron beams. Right: Migrations as present on different levels of event reconstruction, for explanation see text.}
\label{fig:ambig_rec}
\end{center}
\end{figure}

The described scenarios are encountered as shown in Figure~\ref{fig:ambig_rec}. First, for right handed electron beams and therefore right-handed $\tpq$-quarks the polar angle of the $\tpq$-quark $\cos{\mathrm{\theta_{\tpq}}}$ can be reconstructed without any ambiguity. The reconstructed polar-angle spectrum, denoted as {\em Reconstructed} in Fig.~\ref{fig:ambig_rec} agrees very well with the true one, denoted as {\em Monte Carlo}. On the other hand the reconstruction of $\cos{\mathrm{\theta_{\tpq}}}$ in case of left handed $\tpq$-quarks suffers from considerable migrations. Migrations can be caused by either the described physics effect or imprecise event reconstruction. Therefore the event reconstruction has been repeated using only the final state leptons and quarks. The result is called {\em Parton limit} in Fig.~\ref{fig:ambig_rec} and it is visible that the reconstructed  $\cos{\mathrm{\theta_{\tpq}}}$ spectrum deviates from the 'true' polar angle spectrum. This is the consequence of the described ambiguity due to the V-A-structure of the Standard Model. However, a considerable part of the migrations is to be attributed to the imperfect detector reconstruction as can be told from the difference between the 'Monte Carlo' or also 'Parton limit' spectra and the reconstructed spectrum. In a next step the requirements on the quality of the event reconstruction are tightened further. For this Equation~\ref{eq:qual} is applied and only events which satisfy 
\begin{equation}
d^2 < 36
\end{equation}
are accepted. 
After this additional selection criterion the reconstructed spectrum, denoted as {\em Reconstructed with cut} in Fig.~\ref{fig:ambig_rec}, is already much closer to the 'Parton limit'. The additional cut reduces the selection efficiency however to 
52\% where the efficiency for left-handed top quarks is reduced to 43\%.
Finally, the forward-backward asymmetry can be determined with a statistical precision of
\begin{eqnarray}
\mathrm{ \Delta A^{t}_{FB,R}/ A^{t}_{FB,R}} = 1.2\%\,\mathrm{(stat.)} & \textrm{for right-handed $\tpq$-quarks, and}\\
\mathrm{ \Delta A^{t}_{FB,L}/ A^{t}_{FB,L}} = 1.4\%\,\mathrm{(stat.)} &  \textrm{for left-handed $\tpq$-quarks}
\end{eqnarray}
With this statistical precision and that obtained for $\alr$ given in Equation~\ref{eq:precalr} the couplings $\mathrm{Q_L}$ and $\mathrm{Q_R}$ can be measured to a precision of 1\% and and 2\% respectively. This precision allows to distinguish among numerous models~\cite{doublet}. New physics can be probed up to 2.8\,TeV in case of the absence of a $\Zzero '$-boson or up to even 25\,TeV in case a $\Zzero '$-boson exists.

\subsection{Concluding remarks on experimental implications}
The migrations in reconstruction of the $\tpq$-quark direction can be controlled to a 20\% level when imposing tight requirements on the reconstruction quality of the $\tpq$-quark This on the other hand challenges the jet energy resolution of the detector. 
In order to correct for the migration, however, a perfect description of the data by Monte-Carlo simulations of the detector will be required.  The ambiguity could be resolved to a large extent in case the charge of the $\bottom$-quarks could be measured. Both, the control of the energy resolution but in particular the determination of the charge of the $\bottom$-quark may serve as guidelines for the optimisation of the detectors for a future linear collider. The authors of ~\cite{dpn} point out directions on how the charge of the
$\bottom$-quark could be determined.

\section{Summary and outlook}

This article summarises an initial study with full detector simulation to demonstrate the measurement of asymmetries in $\tpq$-quark production at the ILC operated at a centre-of-mass energy of $\roots$=500\,GeV. It is shown that the measurement bares huge potential for the discovery of new physics in the process $\epem \rightarrow \ttbar$ which is very well accessible at the ILC and would contribute considerably to the physics program. 
The results of the study show however that in order to minimise the uncertainties in the measurement tight constraints
are put on the detector precision. The jet energy resolution has to be very well controlled in order to minimise migrations in case of the production of $\tpq$-quarks with left-handed helicity. These migrations would however be best controlled of the charge of the $\bottom$-quark could be measured. Future studies should and will entail an investigation how this can be achieved.
The process investigated here is a benchmark process for the {\em D}etector {\em B}aseline {\em D}ensign, DBD,  of the ILC detectors. This document will be published together with the {\em T}echnical {\em D}ensign {\em R}sport, TDR, of the accelerator at the end of 2012. Both documents will appear at a time when the existence of a Higgs-boson will be confirmed or ruled out by the LHC experiments. In case of the existence of a light Higgs, $\tpq$-quark physics could constitute the second cornerstone of a future linear collider operated at an energy of about $\roots$=500\,GeV. Finally, note that studies have been started on $\tpq$-quark asymmetries at $\roots$=1\,TeV, which will support the physics case for an upgrade of the ILC to this energy.

\section{Acknowledgments}
The authors of the study would like to thank the organisers of the LCWS11 for the opportunity to present the study at the workshop and for the workshop-organisation and the hospitality.
In addition we acknowledge the numerous technical support provided by the ILC software group.



\section{Bibliography}



\begin{footnotesize}


\end{footnotesize}


\end{document}

%% file: symbols.tex
\newcommand{\eplus}{\mathrm{e}^+}
\newcommand{\eminus}{\mathrm{e}^-}
\newcommand{\epem}{\eplus\eminus}

\newcommand{\Zzero}{\mathrm{Z}}

\newcommand{\Wboson}{\mathrm{W}}

\newcommand{\quark}{\mathrm{q}}
\newcommand{\bottom}{\mathrm{b}}
\newcommand{\tpq}{\mathrm{t}}
\newcommand{\afb}{\mathrm{A_{FB}}}
\newcommand{\alr}{\mathrm{A_{LR}}}
\newcommand{\qq}{\mathrm{q}\overline{\mathrm{q}}}

\newcommand{\ttbar}{\mathrm{t}\overline{\mathrm{t}}}

\newcommand{\roots}{\mathrm{\sqrt{s}}}
\newcommand{\invfb}{\mathrm{fb^{-1}}}